\documentclass[journal]{IEEEtran}
\usepackage{cite}{\tiny }
\usepackage{threeparttable}
\usepackage{graphicx}
\usepackage{picinpar}
\usepackage[cmex10]{amsmath}
\usepackage{amsmath,amsfonts,amssymb}
\usepackage{algorithmic}
\usepackage{subfig}
\usepackage{algorithm,setspace}
\usepackage{threeparttable}
\usepackage{caption}
\usepackage{booktabs}
\usepackage{multirow}
\usepackage{diagbox}
\usepackage{etoolbox}
\usepackage{makecell}
\usepackage{amssymb}
\usepackage{mathrsfs}
\usepackage{color}
\definecolor{purple}{RGB}{160,32,240}
\definecolor{darkred}{RGB}{255,0,255}
\usepackage{stfloats}
\usepackage{bm}
\usepackage{multirow}
\usepackage{bbding} 

\begin{document}
	\title{Transformer-Empowered 6G Intelligent Networks: From Massive MIMO Processing to Semantic Communication}
	\author{Yang Wang, Zhen Gao, Dezhi Zheng, Sheng Chen, \IEEEmembership{Fellow, IEEE}, \\Deniz Gündüz, \IEEEmembership{Fellow, IEEE}, and H. Vincent Poor, \IEEEmembership{Life Fellow, IEEE}
	\thanks{Yang Wang, Zhen Gao (corresponding author), and Dezhi Zheng are with Beijing Institute of Technology; Sheng Chen is with the University of Southampton; Deniz Gündüz is with Imperial College London; H. Vincent Poor is with Princeton University.}
}
	\maketitle
	
	\begin{abstract}
		It is anticipated that 6G wireless networks will accelerate the convergence of the physical and cyber worlds and enable a paradigm-shift in the way we deploy and exploit communication networks. Machine learning, in particular deep learning (DL), is expected to be one of the key technological enablers of 6G by offering a new paradigm for the design and optimization of networks with a high level of intelligence. In this article, we introduce an emerging DL architecture, known as the \textit{transformer}, and discuss its potential impact on 6G network design. We first discuss the differences between the transformer and classical DL architectures, and emphasize the transformer's self-attention mechanism and strong representation capabilities, which make it particularly appealing for tackling various challenges in wireless network design. Specifically, we propose transformer-based solutions for various massive multiple-input multiple-output (MIMO) and semantic communication problems, and show their superiority compared to other architectures. Finally, we discuss key challenges and open issues in transformer-based solutions, and identify future research directions for their deployment in intelligent 6G networks.
		
	\end{abstract}
	
	\section{Introduction}\label{S1}
	
	The sixth generation (6G) of wireless cellular networks are expected to connect the cyber and physical worlds, allowing humans to seamlessly interact with a variety of devices in a mixed reality metaverse through connected intelligence. These new and fascinating applications impose challenging requirements and constraints on communication networks, including ultra-high reliability, ultra-low latency, extremely high data rate, substantially high energy and spectral efficiency, ultra-dense connectivity, and a high level of intelligence. These stringent demands of 6G have driven researchers to look for sophisticated physical layer techniques that would go beyond the cycle of incremental improvements. Current wireless networks have been largely designed as a combination of dedicated processing blocks, such as channel estimation, equalization, coding/decoding blocks, where each block is designed separately on the basis of mathematical models that define the statistical behavior of the wireless channels and the underlying data traffic. 
	However, this model-driven and block-based design approach is facing increasing challenges in the complex and diversified scenarios in which 6G networks are expected to operate. The anticipated diversity of devices and hardware technologies, increasing co-existence requirements, and variety of traffic and service demands make such a model-driven approach difficult and inaccurate. 
	In addition, with the deployment of ultra-massive multiple-input multiple-output (MIMO) systems, the optimization of physical layer functionalities based on mathematical models and solutions will become prohibitive due to the computational complexity and associated control overhead. Therefore, it is anticipated that conventional mathematical models and solutions will not be able to provide the required dramatic enhancement in the capacity and performance of future wireless networks.
	
	Recently, machine learning, in particular deep learning (DL), has emerged as a powerful alternative for the design and optimization of wireless networks by learning underlying statistical structure from data instead of building and employing accurate mathematical models \cite{DL_survey0}. The potential impact of DL-based solutions has already been shown in a variety of challenging wireless communication problems, in which it is either difficult to obtain a model of the system, or the complexity of the model does not lend itself to tractable solutions with feasible computational complexity \cite{DL_survey0, DL_survey1}. 
	
	While DL-based solutions are appealing, the actual deployment is still challenging as they require architecture and hyperparameter optimization for each specific task. Therefore, proposing a more efficient and widely applicable DL architecture is essential for solving complex communication problems. A novel deep neural network (DNN) structure, called the \textit{transformer}, has emerged recently, and achieved remarkable success in a variety of natural language processing (NLP) and computer vision (CV) tasks \cite{Transformer}. The transformer architecture is built upon the \textit{self-attention} mechanism, which relates different parts of a data sequence for a more accurate representation of the sequence. Self-attention layers in the transformer architecture enable a global receptive field, and the multi-head mechanism ensures that the network can pay attention to multiple discriminative parts of the inputs. By highlighting the transformer's multi-model fusion and feature representation capabilities, we explore its application in 6G intelligent network design, and propose a new transformer-based intelligent processing architecture. We focus on massive MIMO and semantic communication applications; however, we expect the transformers to find applications in many other components of future data-driven 6G networks. 
	
	The rest of the article is organized as follows. The following section briefly introduces the application of DL in wireless communications. Then, we introduce the transformer architecture. Next, we present a transformer-based architecture for 6G intelligent processing, and study its performance in various wireless communication problems. We then discuss open research issues in transformer-empowered 6G intelligent networks and conclude the article.
	
	
	\begin{figure}[tp!]
		\vspace*{-2mm}
		\begin{center}
			\includegraphics[width=\columnwidth, keepaspectratio]{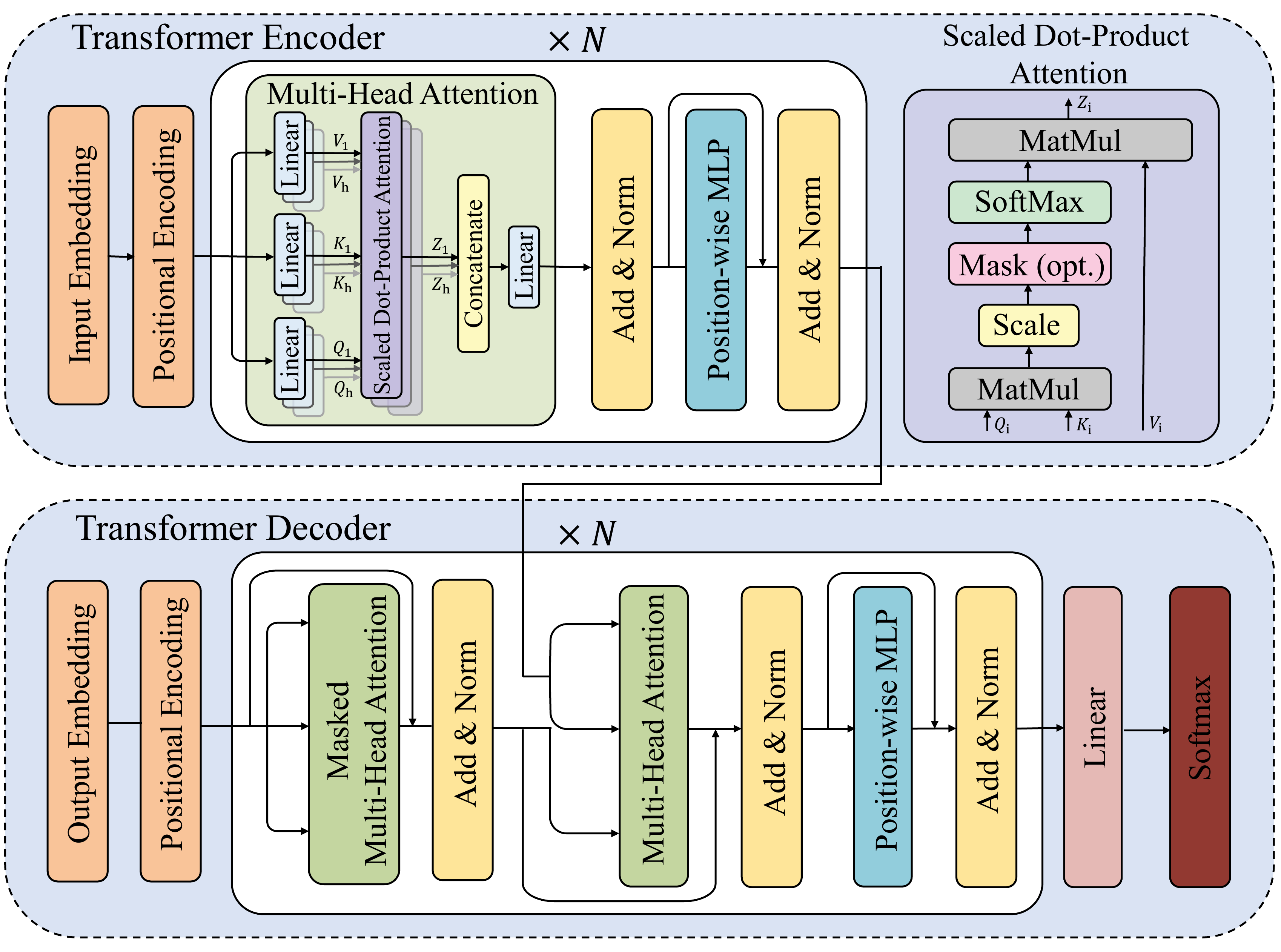}
		\end{center}
		\vspace*{-4mm}
		\captionsetup{font={footnotesize}, name={Fig.},labelsep=period}
		\caption{Structure of the transformer network.}
		\label{net_framework} 
		\vspace*{-3mm}
	\end{figure}
	
	\section{Overview of Deep Learning and the Transformer Architecture}\label{S2}
	
	DL is a powerful computational tool for understanding complex data representations and patterns, and as such, offers a new paradigm to tackle complicated problems in communication network design. In this section, we briefly provide some background on popular DNN architectures and their applications in wireless communications.
	
	\subsection{Common DNN Architectures}\label{S2.1}
	
	Classic neural network architectures include multi-layer perception (MLP), convolutional neural network (CNN), recurrent neural network (RNN), and stacked autoencoder (SAE).
	
	MLP is an artificial neural network that consists of at least three layers of fully-connected neurons, parameterized by a substantial number of connection weights. Under the premise of keeping the same input and output dimensions, the computational complexity of the fully-connected layer is given by $\mathcal{O}(n^2\cdot d^2)$\footnote{Note that, a two-dimensional sequence $\mathbf{X}\in\mathbb{R}^{n\times d}$ is used to analyze the complexity of different DNN structures, where $n$ is the sequence length, and $d$ is the representation dimension.}, where the input vector ${\bf x}\in\mathbb{R}^{1\times nd}$ is reshaped from the two-dimensional sequence $\mathbf{X}\in\mathbb{R}^{n\times d}$. MLP-based solutions have been developed to address various wireless communication problems, such as channel estimation and beamforming \cite{DL_survey1}. It has been observed that deeper networks typically provide better generalization; however, training fully-connected deep networks suffers from high complexity and low convergence efficiency.
	
	To reduce the training complexity, CNNs employ a set of locally connected kernels, rather than fully-connected layers, to capture local correlations between different data regions. Compared with MLP, CNN reduces the number of model parameters significantly and maintains the affine invariance by leveraging three important ideas: sparse interactions, parameter sharing, and equivariant representations. The computational complexity for the convolutional layer is given by $\mathcal{O}(k\cdot n\cdot d^2)$, where $k\times d$ is the kernel size to adapt sequential processing\cite{Transformer}. By treating the channel matrices as two-dimensional images, CNNs have shown great potential for tasks such as channel estimation, channel state information (CSI) feedback, beamforming\cite{DL_survey1}, as well as semantic image transmission~\cite{SC-CNN}.
	
	RNNs constitute another class of DNN architectures that exploit sequential correlations between samples. At each step, it produces the output via recurrent connections between hidden units. However, the traditional RNN architecture is slow to train, and suffers from vanishing and exploding gradients. Long short-term memory (LSTM) architecture mitigates these problems by introducing a set of gates, which allow memory to be restored across longer sequences. The computational complexity for the recurrent layer is given by $\mathcal{O}(n\cdot d^2)$ \cite{Transformer}. Recently, there have been several works utilizing LSTMs to extract temporal correlations across data, (e.g., in channels with memory) for communication system design\cite{DL_survey1}.
	
	SAE architecture consists of hierarchically connected multiple autoencoders. Its basic component, autoencoder, contains two parts: an encoder that acquires a low-dimensional representation of input, and a decoder that reconstructs the input from the compressed vector. SAE is widely used to extract features and patterns that contain essential and compressed information about data. From a learning perspective, the entire communication system can be viewed as an end-to-end SAE, and its multiple sub-modules can also be viewed as SAEs, including pilot design and channel estimation, CSI feedback, and semantic communications~\cite{DL_survey1,SC-CNN}. Thus, SAE is a core DNN structure for many of the current DL-based communication system components.
	
	\subsection{Self-Attention and Transformer}\label{S2.2}	
	
	Although MLP, CNN, RNN, and SAE have been widely utilized in DL-based communication system design with some success, efforts continue to push the boundaries of DL models in practical communication systems. Recently, the evolution of DNN architectures in NLP has led to a prevalent architecture known as the transformer \cite{Transformer}. We argue that the transformer holds a great potential also in the design of intelligent communication systems.
	
	As shown in Fig.~\ref{net_framework}, the transformer is a sequence-to-sequence DNN model and consists of an encoder and a decoder module with several encoder/decoder layers of the same architecture. The input and output sequences are converted to vectors of dimension $d$ by embedding and positional encoding layers. Each encoder/decoder layer has the same structure, and is mainly composed of a self-attention sub-layer following by a position-wise MLP sub-layer, while each decoder also contains a masked attention sub-layer before the self-attention sub-layer. For building a deep model, a residual connection is employed around each sub-layer, followed by a layer normalization module.
	
	
	\begin{figure}[tp!]
		\vspace*{-2mm}
		\begin{center}
			\includegraphics[width=\columnwidth, keepaspectratio]{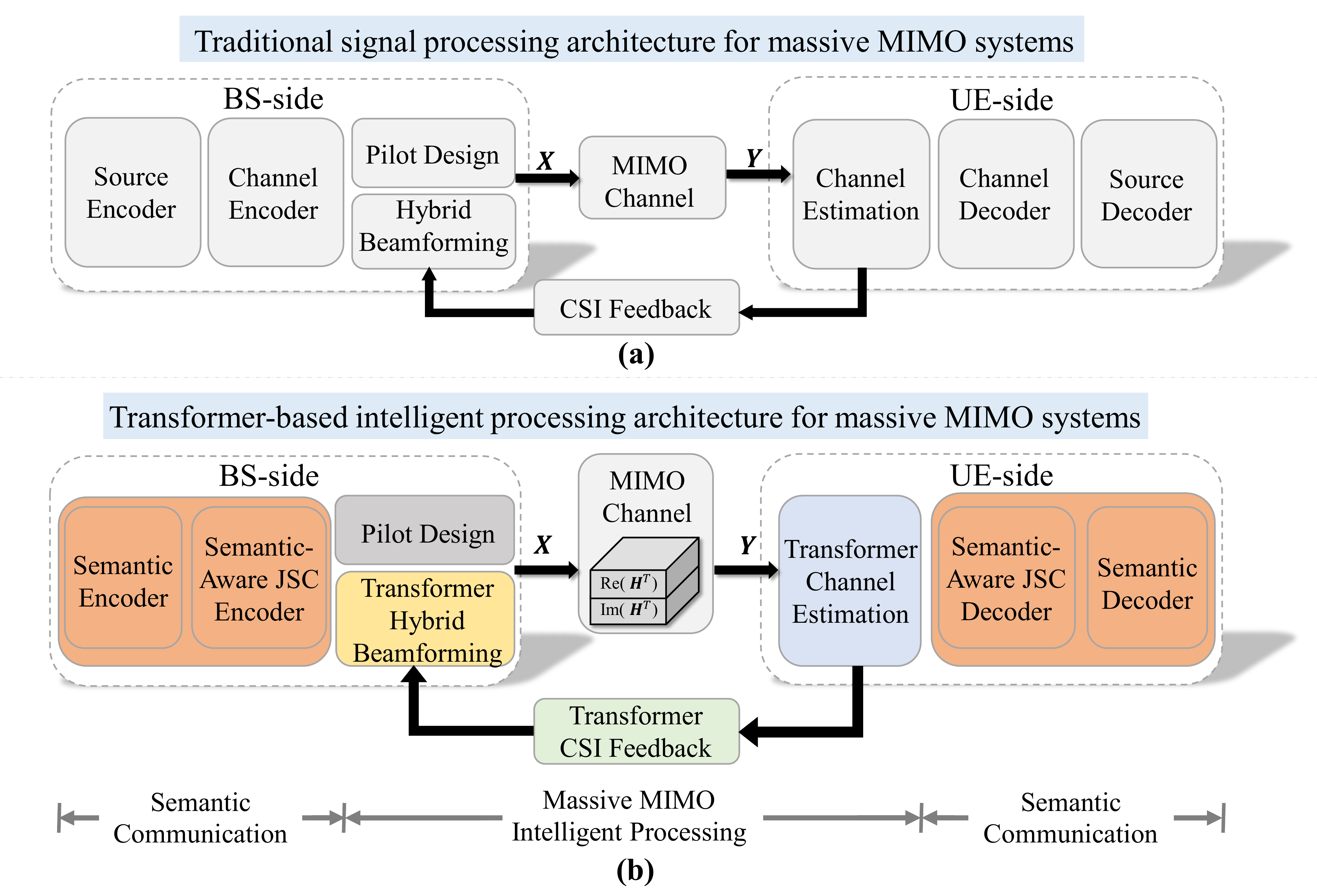}
		\end{center}
		\vspace*{-4mm}
		\captionsetup{font={footnotesize}, name={Fig.},labelsep=period}
		\caption{Traditional and proposed transformer-based signal processing architecture for massive MIMO systems.}
		\label{MIMO_framework} 
		\vspace*{-4mm}
	\end{figure}
	
	Self-attention mechanism relates different positions in a single sequence to compute a representation of the sequence, which can also be regarded as a non-local filtering operation.
	In a single-head self-attention layer, the input sequence $\mathbf{X}\in\mathbb{R}^{n\times d}$ is first transformed into three different sequential vectors: the query $\mathbf{Q}\in\mathbb{R}^{n\times d_{k}}$, the key $\mathbf{K}\in\mathbb{R}^{n\times d_{k}}$ and the value $\mathbf{V}\in\mathbb{R}^{n\times d_{v}}$ by three different linear matrices, which are obtained through training. Here, $d_{k}$ and $d_{v}$ are the dimensions of query (key), and value subspaces, respectively. Subsequently, as shown in Fig.~\ref{net_framework}, the scale dot-production attention operation generates the attention weights by aggregating the query and the corresponding key. The resulting weights are assigned to the corresponding value, yielding the output vectors. To facilitate the complexity analysis, we assume that query, key, and value matrices have the same dimension as the input sequence, i.e., $d_k=d_v=d$. Thus, the complexity of self-attention layer can be expressed as $\mathcal{O}(n^2 \cdot d)$ \cite{Transformer}. In terms of computational complexity, self-attention layers are significantly faster than fully-connected layers, and are faster than recurrent layers when the sequence length $n$ is smaller than the representation dimensionality $d$. The training efficiency of recurrent layers is much lower than that of the self-attention layers due to the sequential processing. Furthermore, since convolutional layers are generally more complex than recurrent layers, by a factor of $k$, their complexity is also higher than the self-attentive layer.
	Instead of performing single-head self-attention with query, key, and value, multi-head attention allows the model to jointly attend to information from different representation subspaces at different positions. Specifically, different heads use different three group linear matrices, and these matrices can project the input vectors into multiple feature subspaces (i.e., ${\{\mathbf{Q}_i\}}_{i=1}^h$, ${\{\mathbf{K}_i\}}_{i=1}^h$, and ${\{\mathbf{V}_i\}}_{i=1}^h$, where $h$ is the number of heads) and processes them by several parallel attention heads (layers). The resulting vectors are concatenated and mapped to the final output. 
	
	The position-wise MLP sub-layer is a fully-connected feed-forward module that operates separately and identically on each position. This module consists of two linear transformations with ReLU activation, where the parameters are shared across different positions, and the complexity is $\mathcal{O}(n \cdot d^2)$ \cite{Transformer}. Since the transformer does not introduce recurrence or convolution, it has no knowledge of positional information (especially for the encoder). Thus, additional positional information is introduced through positional encoding in order to model the relative positions of the input sequences.
	
	Compared with CNN/RNN models, the transformer makes few assumptions about the underlying structure of data, which makes it a universal and flexible architecture.
	The non-sequential nature of the transformer architecture allows it to capture long-range dependencies in the input data through self-attention. Not surprisingly transformers have also shown remarkable success in semantic communications \cite{SC_DL_1} for the transmission of text over noisy channels. In this article, we show that transformers can have a critical role in other communication tasks as well.
	
	\begin{figure*}[tp!]
		\vspace*{-3mm}
		\begin{center}
			\subfloat[]{
				\label{CE_frame}
				\includegraphics[width=.66\textwidth,keepaspectratio]{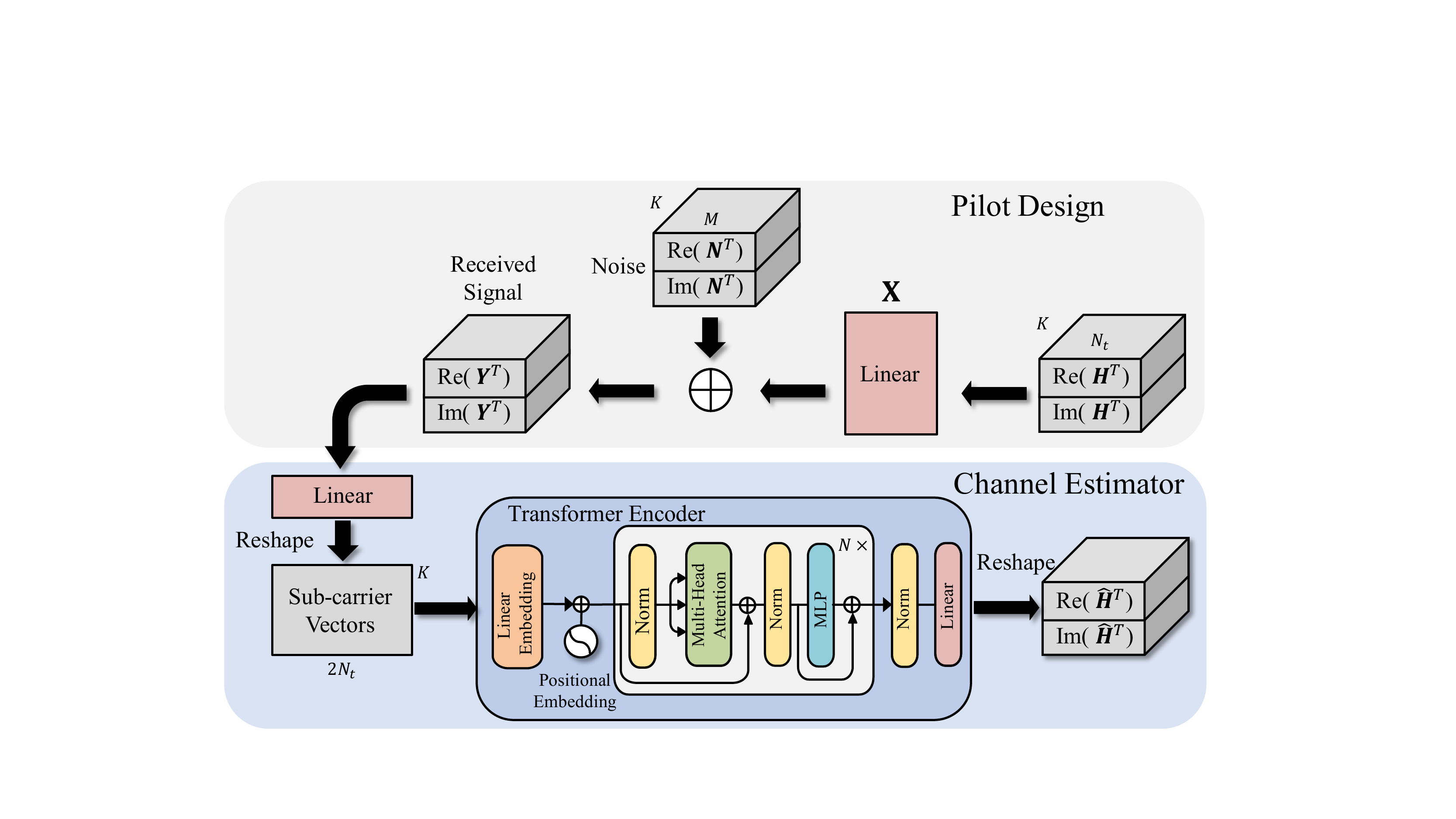}
			}
			\subfloat[]{
				\label{CE_Per}
				\includegraphics[width=.33\textwidth,keepaspectratio]{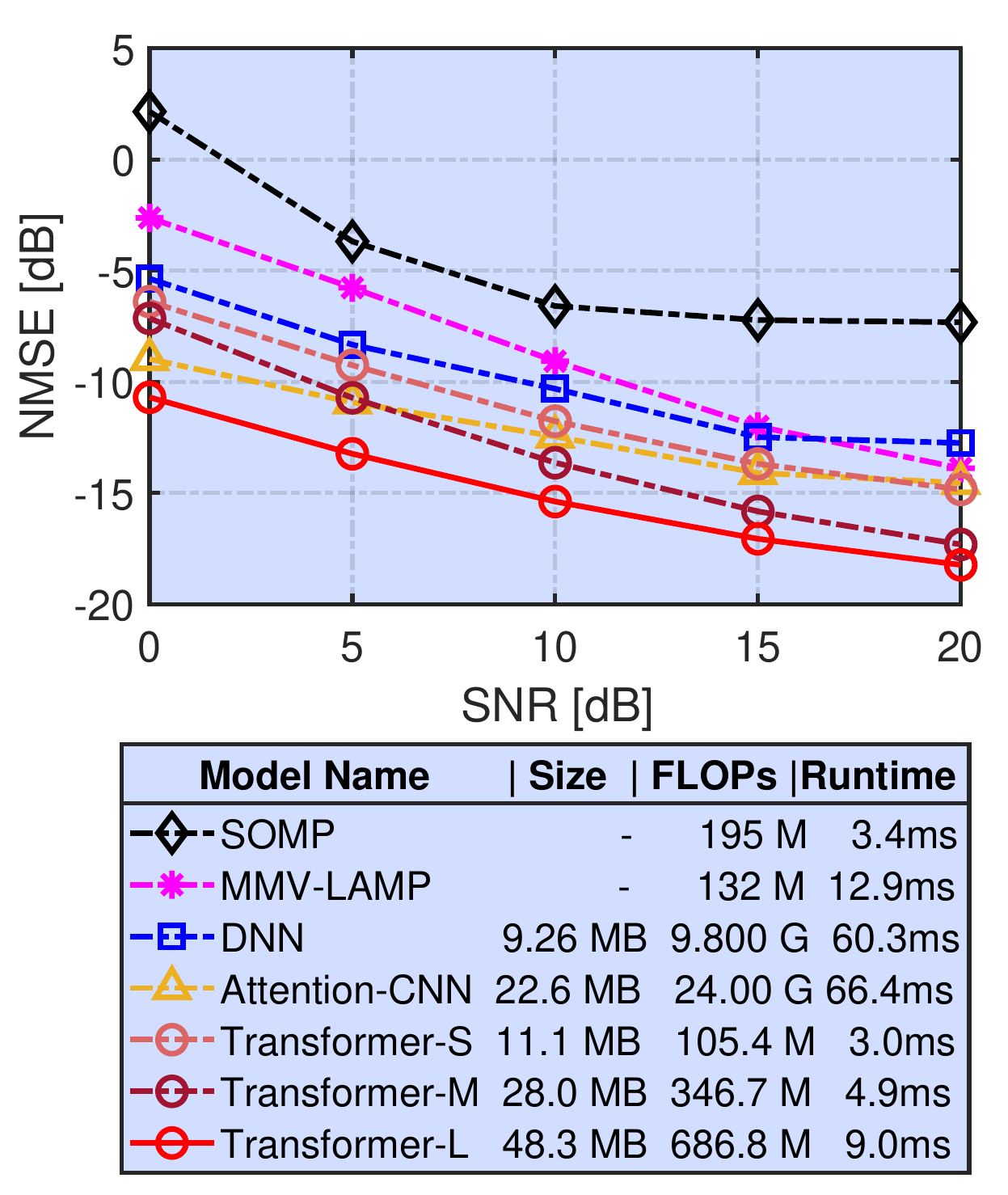}
			}
		\end{center}
		\vspace*{-4mm}
		\captionsetup{font={footnotesize}, name={Fig.},labelsep=period}
		\caption{(a)~The transformer-based end-to-end architecture for jointly designing the pilot signals and channel estimator; and (b)~NMSE performance comparison of different channel estimation schemes vs. signal-to-noise ratio (SNR).}
		\label{CE} 
		\vspace*{-4mm}
	\end{figure*}
	
	\section{Transformer For 6G Intelligent Processing}\label{S3}
	
	Massive MIMO is an essential physical layer technology to accommodate the exponential growth of mobile data traffic. Fig.~\ref{MIMO_framework}\,(a) illustrates a generic communication system, divided into two parts: the MIMO processing part and source \& channel coding part. The former includes pilot design, channel estimation, CSI feedback, and hybrid beamforming (HBF). The latter is composed of source coding and channel coding. We seek to expand the applicability of the transformer to serve as a general-purpose backbone for these crucial modules. In particular, as illustrated in Fig.~\ref{MIMO_framework}\,(b), we propose a novel 6G intelligent processing architecture employing transformer for both the massive MIMO intelligent processing blocks and the newly emerging  semantic communication blocks.
	
	\subsection{Channel Estimation}\label{S3.1} 
	
	Accurate CSI at the base station (BS) is critical for beamforming and signal detection in massive MIMO systems. However, CSI acquisition overhead of conventional orthogonal pilot-based approaches increases linearly with the number of antennas. To reduce the pilot overhead, existing 5G NR standard limits the number of pilots to be significantly smaller than the number of antennas. However, it is challenging to accurately estimate high-dimensional channels with few pilots. By exploiting the sparsity of the channels in the angular and/or delay domains, compressive sensing (CS)-based solutions have been proposed to overcome this issue. Nevertheless, since the dimension of the CSI to be estimated is extremely large, the involved matrix inversion operations and the iterative nature of CS-based techniques result in prohibitively high computational complexity and storage requirements.	
	
	More recently, researchers have resorted to DL techniques to overcome the aforementioned challenges. A learned approximate message passing (LAMP) network is proposed in \cite{LAMP} to mitigate the performance degradation of the original AMP algorithm, since a priori model may not always be consistent with the actual system. Further, by exploiting the channels’ structured sparsity, an improved multiple-measurement-vector LAMP (MMV-LAMP) network \cite{MMV-LAMP} can jointly recover multiple subcarriers' channel conditions with improved accuracy.
	The authors of \cite{CE-DNN} propose an end-to-end DNN architecture to jointly design the pilot signals and channel estimator. A CNN module combined with non-local attention layer is employed in \cite{attention-CNN} to exploit longer range correlations in the channel matrix.
	
	Nevertheless, most existing DL-based channel estimation solutions are based on the MLP and CNN architectures. Here, we propose a novel channel estimator that utilizes the universal and flexible transformer architecture, as illustrated in Fig.~\ref{CE}\,(a). Specifically, the proposed transformer-based solution includes a dimensionality reduction network for pilot design and a reconstruction network for channel estimation. 
	We exploit a fully-connected linear layer to learn the pilot sequences\cite{MMV-LAMP,attention-CNN,CE-DNN}. More importantly, in our channel estimation module, the encoder part of the transformer is exploited to reconstruct the channel. Unlike local-attention in\cite{attention-CNN}, self-attention in the transformer can extract long-range correlations between subcarriers and adjust the weight of each subcarrier, so that the global features of the channel matrix can be extracted for enhanced estimation accuracy.
	
	To evaluate the performance of the proposed transformer-based solution, we investigate the downlink channel estimation problem in $M$ successive time slots, where the BS is equipped with a uniform planar array (UPA) with $N_t = 8\times 8=64$ antennas, the user equipment (UE) has single-antenna, the number of orthogonal frequency division multiplexing (OFDM) sub-carriers is $K=32$, and the channel estimation compression ratio is $\rho=\frac{M}{N_t}=\frac{3}{8}$. We consider a sparse channel scenario with $N_c=6$ clusters, $N_p=10$ paths per cluster, and an angle spread of $\Delta\theta=\pm 3.75^\circ$. We generate training, validation, and test datasets of 100,000, 10,000, 5,000 samples, respectively. We consider the normalized mean square error (NMSE) as the performance metric. 
	
	To illustrate the advantages of our proposed channel estimator in Fig.~\ref{CE}\,(a), we compare it with four benchmarks. The first one is the traditional simultaneous orthogonal matching pursuit (SOMP) based estimator, denoted as `SOMP'. The second and third are the conventional DL-based channel estimators, namely, the MMV-LAMP based estimator \cite{MMV-LAMP} and the DNN-based estimator \cite{CE-DNN}, denoted as `MMV-LAMP' and `DNN', respectively. Finally, we consider the state-of-the-art attention-CNN based channel estimator \cite{attention-CNN}, abbreviated as `Attention-CNN', as the fourth benchmark.
	We propose three distinct transformer-based estimators with different model sizes, denoted as `Transformer-S', `Transformer-M', and `Transformer-L', respectively. Fig.~\ref{CE}\,(b) shows the NMSE performance of different channel estimation schemes. Evidently, the proposed transformer-based estimator significantly outperforms the conventional and other DL-based methods, especially with comparable model sizes. We observe that, during the inference stage, the floating-point operations per second (FLOPs) and the runtime per sample of the transformer-based estimator are much lower than those of other DL-based methods. Moreover, we can observe that the performance of the transformer improves with the model size. This demonstrates that the transformer-based method can learn latent features from the data more effectively to achieve better channel estimation accuracy with less pilot overhead. It also provides a flexible trade-off between the model complexity and performance, and the users can choose the operating point based on the underlying resources and application requirements.
	
	\begin{figure*}[tp!]
		\vspace*{-4mm}
		\begin{center}
			\subfloat[]{
				\label{CSI_frame}
				\includegraphics[width=.69\textwidth,keepaspectratio]{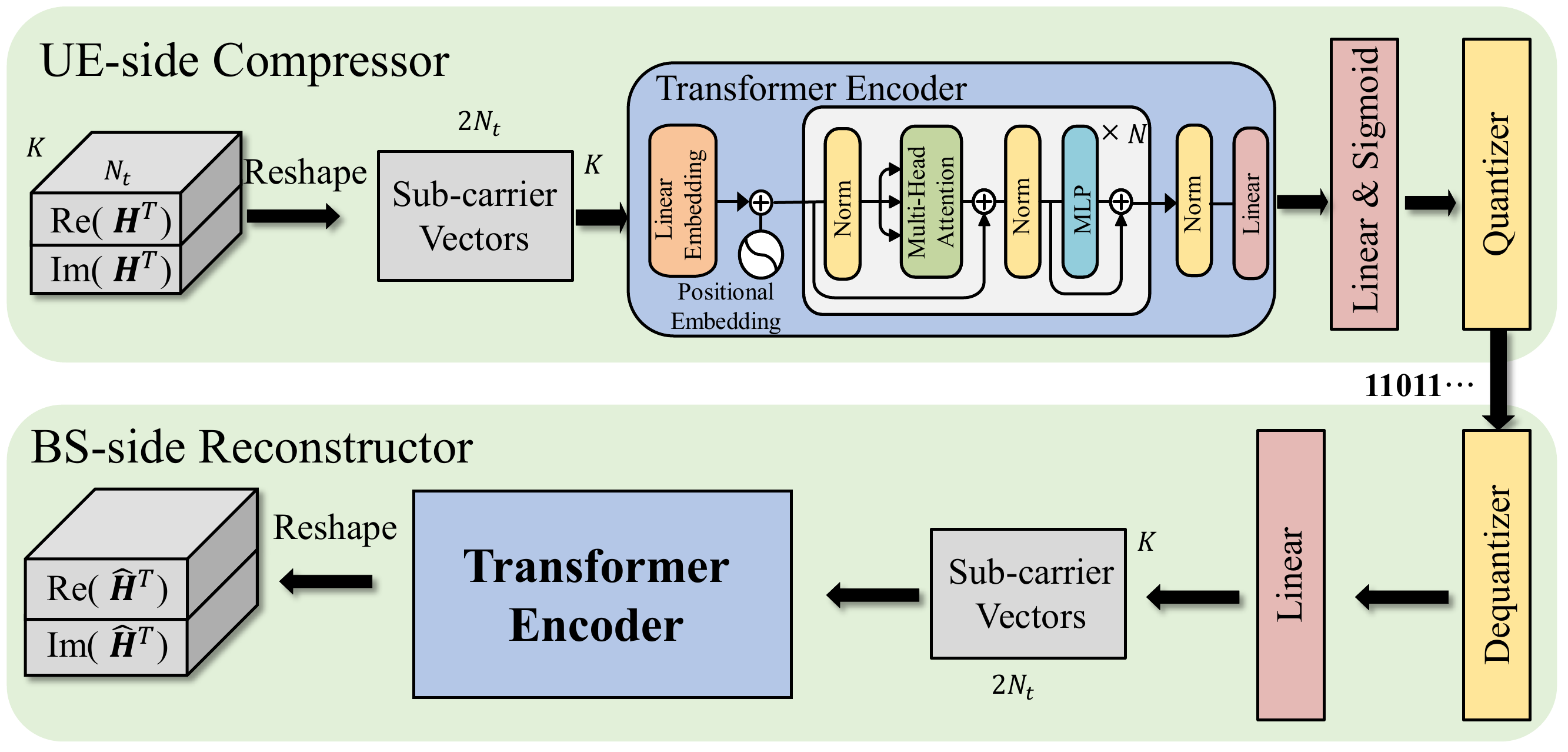}
			}
			\subfloat[]{
				\label{CSI_Per}
				\includegraphics[width=.29\textwidth,keepaspectratio]{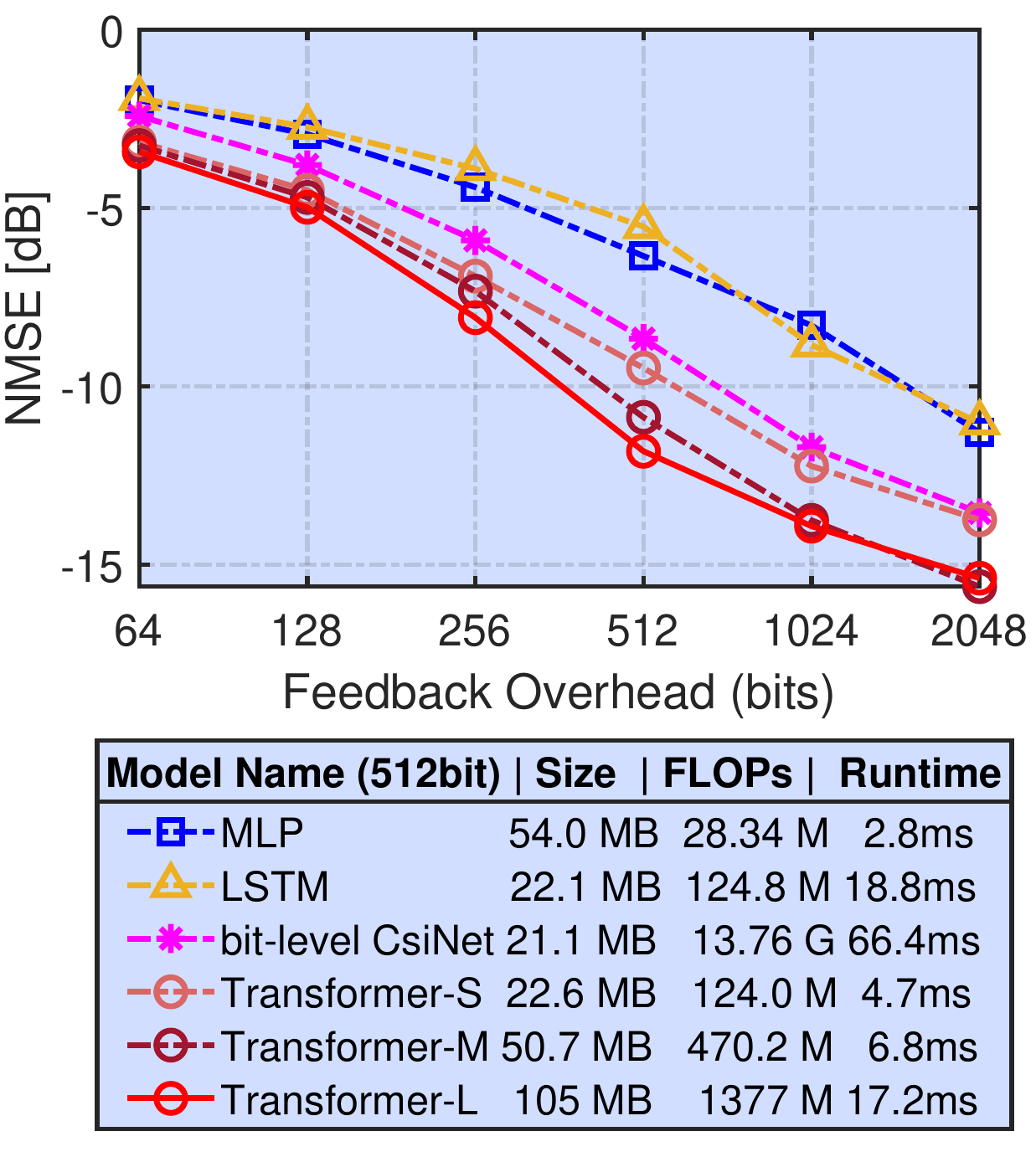}
			}
		\end{center}
		\vspace*{-3mm}
		\captionsetup{font={footnotesize}, name={Fig.},labelsep=period}
		\caption{(a)~The transformer-based CSI feedback architecture; and (b)~NMSE performance comparison of different CSI feedback schemes vs. feedback overhead.}
		\label{CSI-feedback} 
		\vspace*{-3mm}
	\end{figure*}
	
	\subsection{CSI Feedback}\label{S3.2}
	
	CSI feedback is essential in frequency-division duplex (FDD) systems. For time-division duplex (TDD) systems, by exploiting channel reciprocity, the transmitter may estimate the downlink CSI from the uplink CSI. But such reciprocity relies on many ideal factors, including the accurate calibration of the transceiver RF chains at both the BS and UE. For massive MIMO, the perfect uplink and downlink reciprocity is difficult to achieve, and the BS has to rely on CSI feedback for both FDD and TDD operations. However, the large number of antennas result in excessive feedback overhead. Similarly to channel estimation, CS-based techniques can be used to reduce the CSI feedback overhead. However, these techniques cannot fully exploit the channel structure since the channels in real systems are not exactly sparse. Recently, DL-based solutions have achieved impressive results for CSI feedback. An autoencoder architecture, called CsiNet, is proposed in \cite{CsiNet} to reduce the feedback overhead in massive MIMO systems, which is shown to outperform traditional CS-based methods in terms of both compression ratio and recovery accuracy. Moreover, a bit-level CsiNet is designed by considering the effects of CSI quantization distortion \cite{CSI-CNN}. This design can be easily assembled in existing CSI networks with some slight modifications. Subsequent studies expanded and designed various network models based on CNN and LSTM architectures to handle different CSI feedback problems \cite{DL_survey1,CSI-LSTM}.
	
	Herein, we present a transformer-based CSI feedback scheme to obtain more efficient quantization and compression performance compared with the prior work. As illustrated in Fig.~\ref{CSI-feedback}\,(a), we utilize the fully-connected linear layer to linearly embed the channel data and use $\mathrm{sine}$ and $\mathrm{cosine}$ functions of different frequencies to represent the relative positions of the sub-carriers. Then, the transformer encoder extracts features from the channel data embedded with the positional information. Next, the features are vectorized, and a fully-connected linear layer is used to generate a real-valued compressed codeword. The codeword is then converted to the feedback bit-stream through a quantization layer, which is constructed by uniform scalar quantization\cite{CSI-CNN}. Since the whole network structure corresponds to the compression recovery task, the decoder adopts the same structure as the encoder.
	
	We use the same simulation parameters of Subsection~\ref{S3.1} to evaluate the proposed transformer-based CSI feedback schemes with different model sizes, denoted as `Transformer-S', `Transformer-M', and `Transformer-L', respectively. Three benchmark schemes with bit-level feedback are also considered for comparison. The first one is the MLP-based CSI feedback scheme, `MLP', where the encoder and decoder consist of three fully-connected layers, respectively. The second one is the `bit-level CsiNet' scheme in \cite{CSI-CNN}, while the third is the `LSTM' scheme in \cite{CSI-LSTM}. We again use the NMSE metric for performance evaluation. Fig.~\ref{CSI-feedback}\,(b) shows that all three transformer-based CSI feedback schemes outperform the three benchmarks. Meanwhile, the FLOPs of the transformer-based schemes are much lower than `bit-level CsiNet', and all the proposed schemes have lower runtime than both the `bit-level CsiNet' and the `LSTM' schemes. Also, we can observe that the performance of the transformer improves with the model size, providing a trade-off between complexity and performance. We can see that `Transformer-S' is sufficient when a few feedback overhead is desired, while the more complex alternatives provide further gains as the feedback overhead increases. In a nutshell, the transformer can better extract these implicit features in the CSI and fewer feedback bits are needed to reconstruct the CSI at the BS with the same quality, which reduces the feedback overhead and latency.
	
	\begin{figure*}[tp!]
		\vspace*{-4mm}
		\begin{center}
			\hspace*{-1mm}\subfloat[]{
				\label{HBF-frame}
				\includegraphics[width=.68\textwidth,keepaspectratio]{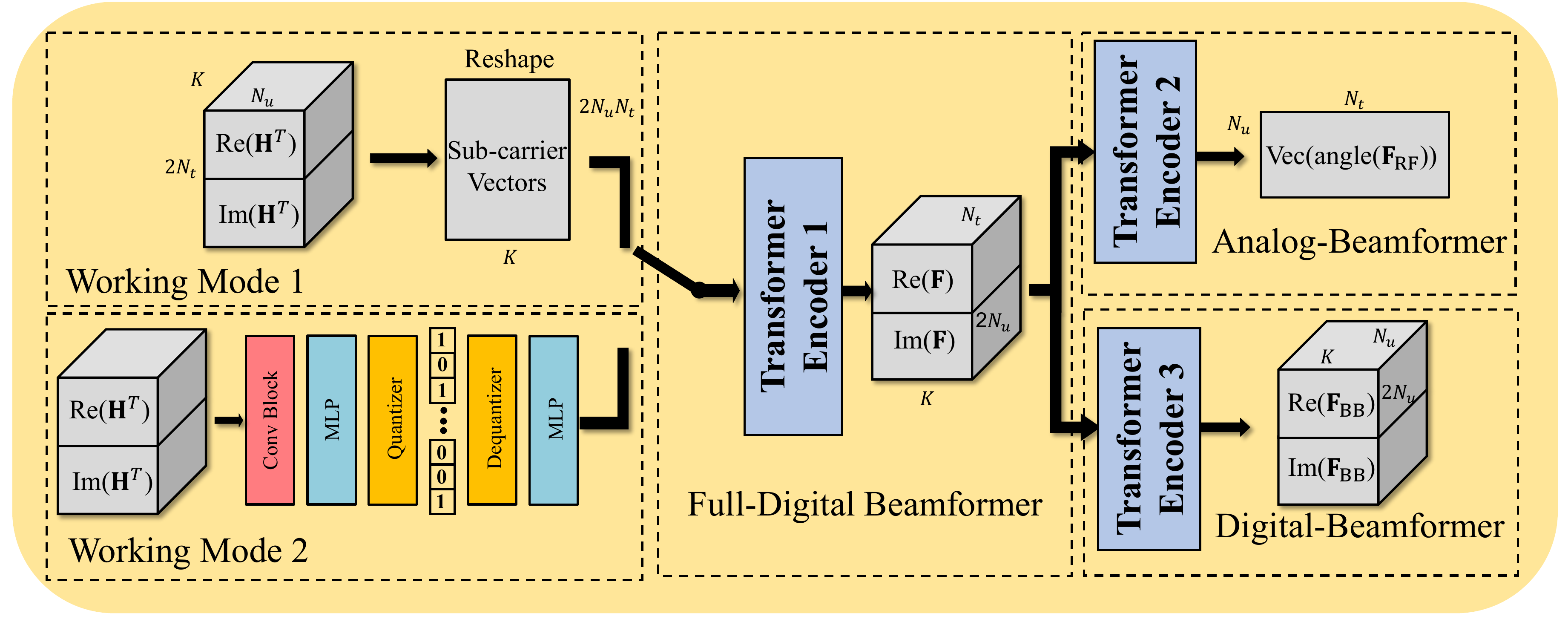}
			}
			\hspace*{-2mm}\subfloat[]{
				\label{HBF_Per}
				\includegraphics[width=.32\textwidth,keepaspectratio]{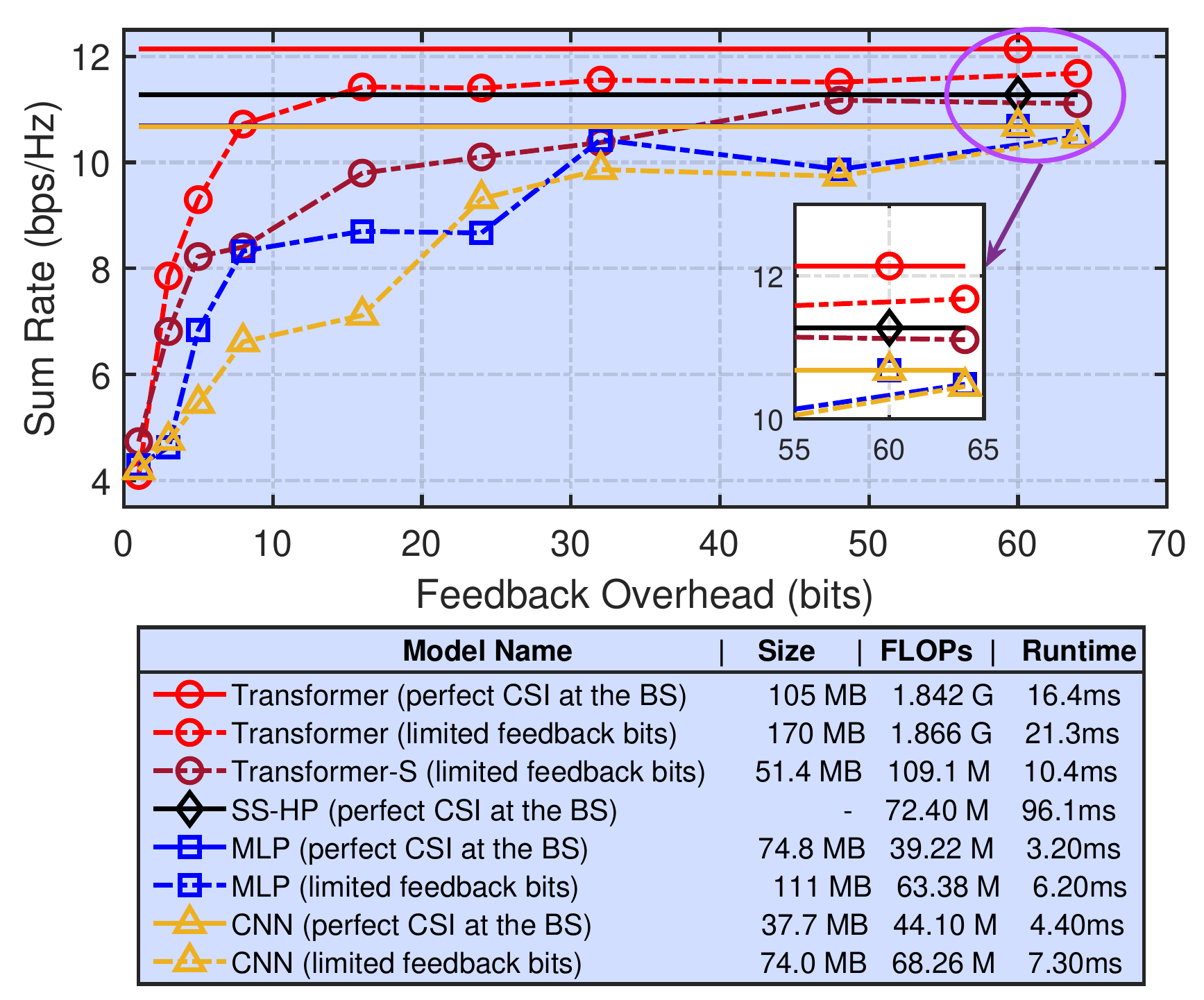}
			}
		\end{center}
		\vspace*{-3mm}
		\captionsetup{font={footnotesize}, name={Fig.},labelsep=period}
		\caption{(a)~The transformer-based HBF architecture; and (b)~Sum-rate vs. feedback overhead for different HBF schemes.}
		\label{HBF} 
		\vspace*{-4mm}
	\end{figure*}	
	
	\subsection{Hybrid Beamforming}\label{S3.3} 
	
	Conventional massive MIMO systems with fully-digital architecture require a dedicated RF chain for each antenna, which results in excessive power consumption and extremely high RF hardware costs. The alternative hybrid analog-digital MIMO architecture employs a much lower number of digital RF chains than the number of antennas, where each RF chain is connected to multiple active antennas, and the signal phase on each antenna is controlled via a network of analog phase shifters. The analog phase shifter can be seen as a low-cost passive device, which controls only the phase of the signal. Compared with its fully-digital counterpart, HBF optimization is significantly more challenging due to the constant modulus constraint on the analog beamformer \cite{HBF_1}. 
	
	Many model-based solutions have been proposed to tackle this challenge. For instance, the authors of \cite{HBF_1} propose spatial sparse hybrid precoding (SS-HP) to achieve near fully-digital performance by exploiting channel sparsity. However, model-based HBF algorithms require time consuming optimization iterations to obtain near-optimal solutions. Moreover, they demand either perfect downlink CSI or a codebook with an accurate sparse basis, which are difficult to acquire in practice. To overcome these issues, DL-inspired beamforming has been proposed, whereby prior information is captured from radio channel measurements. In \cite{HBF-CNN}, the authors propose a CNN-based HBF architecture that can be trained to maximize spectral efficiency with imperfect CSI. The authors of \cite{HBF-MLP} propose a MLP-based downlink multi-user HBF module to maximize the spectral efficiency from limited CSI feedback bits.
	
	To the best of our knowledge, all the existing DL-based HBF schemes adopt the MLP or CNN architectures. We propose a transformer-based HBF scheme, composed of three transformer encoder modules, as shown in Fig.~\ref{HBF}\,(a). According to \cite{HBF_1}, analog RF beamformer and digital baseband beamformer can be optimized to approach the optimal fully-digital beamformer. Motivated by this principle, each transformer encoder in Fig.~\ref{HBF}\,(a) implements a part of the HBF optimization. 
	More specifically, the input dimension of the first transformer encoder is $K\times 2 N_u N_t$, where $N_u$ is the number of UEs, and the output represents the fully-digital beamformer, i.e., ${\bf{F}}\in\mathcal{C}^{K\times 2 N_u N_t}$; the input dimension of the second transformer encoder is the permutation of ${\bf{F}}$, i.e., $N_u\times 2 K N_t$, and the output is the phase of the analog RF beamformer, i.e., ${\bf {P}}={\rm{vec}}({\rm{angle}}({\bf{F}}_{\rm RF}))\in\mathcal{C}^{N_u\times N_t}$; the third transformer encoder represents the digital baseband beamformer, which takes ${\bf{F}}$ as input and produces ${\bf{F}}_{\rm BB}\in\mathcal{C}^{K\times 2N_uN_u}$ as output, respectively.
		By introducing the structural prior information of traditional optimization methods, combined with the self-attention's feature extraction ability, we can achieve better performance than traditional as well as existing DL-based methods in the literature. 
		As shown in Fig.~\ref{HBF}\,(a), we consider two working modes: 1) the first mode requires an estimated CSI matrix as input, which is achieved by the adopted CSI feedback scheme; 2) the second mode relies on implicit CSI as input, which is conveyed by the feedback bits transmitted from the UEs, and in this case, the CSI feedback network is jointly trained with the proposed HBF network. Note that the case in which the proposed HBF network is trained with the perfect CSI matrix as input (working mode 1), can be regarded as an upper bound for the case trained with quantized CSI feedback bits (working mode 2).
	
	To illustrate the superior performance of our transformer-based HBF, we use the channel parameters similar to those in Subsection~\ref{S3.1}. We set the number of UEs to $N_u=2$. We choose three benchmarks for comparison, namely SS-HP from \cite{HBF_1}, CNN-based HBF of \cite{HBF-CNN}, and MLP-based HBF from\cite{HBF-MLP}. 
	The sum rate comparison of different schemes is depicted in Fig.~\ref{HBF}\,(b). It can be seen that the transformer-based HBF scheme significantly outperforms SS-HP and other DL-based HBF schemes with both complete and limited CSI feedback. The performance gains over the benchmarks are particularly considerable at low feedback overhead of 3 to 24\,bits. Moreover, the proposed scheme with limited feedback bits even outperforms SS-HP with perfect CSI, when the feedback overhead is greater than 24\,bits. This demonstrates the effectiveness of the proposed transformer-based HBF architecture, particularly under the practical limited feedback scenario. However, both `Transformer' and `Transformer-S' have higher FLOPs and runtime than other DL-based schemes. Therefore, it is of interest to develop a more efficient transformer-based HBF architecture with guaranteed performance.
	
	\begin{figure*}[tp!]
		\vspace*{-4mm}
		\begin{center}
			\hspace*{-2mm}\subfloat[]{
				\label{SC-frame}
				\includegraphics[width=.67\textwidth,keepaspectratio]{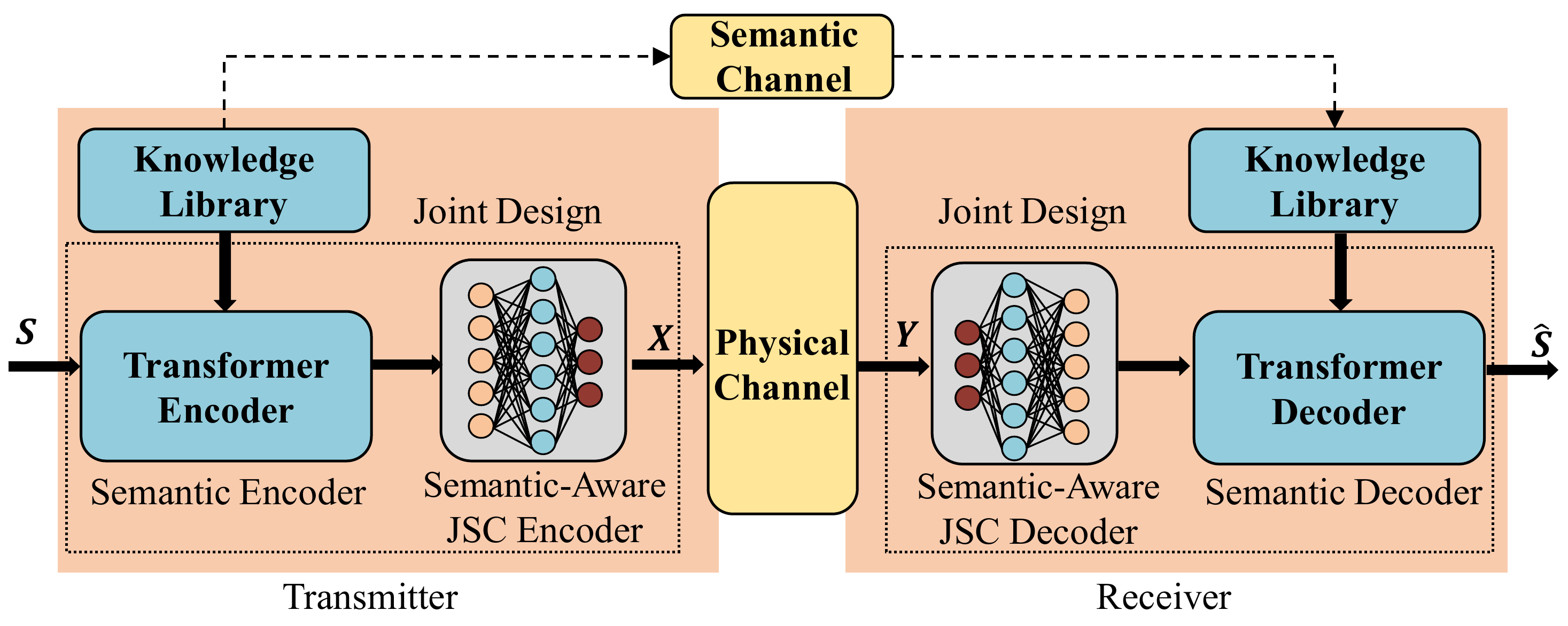}
			}
			\hspace*{-2mm}\subfloat[]{
				\label{SC_Per}
				\includegraphics[width=.33\textwidth,keepaspectratio]{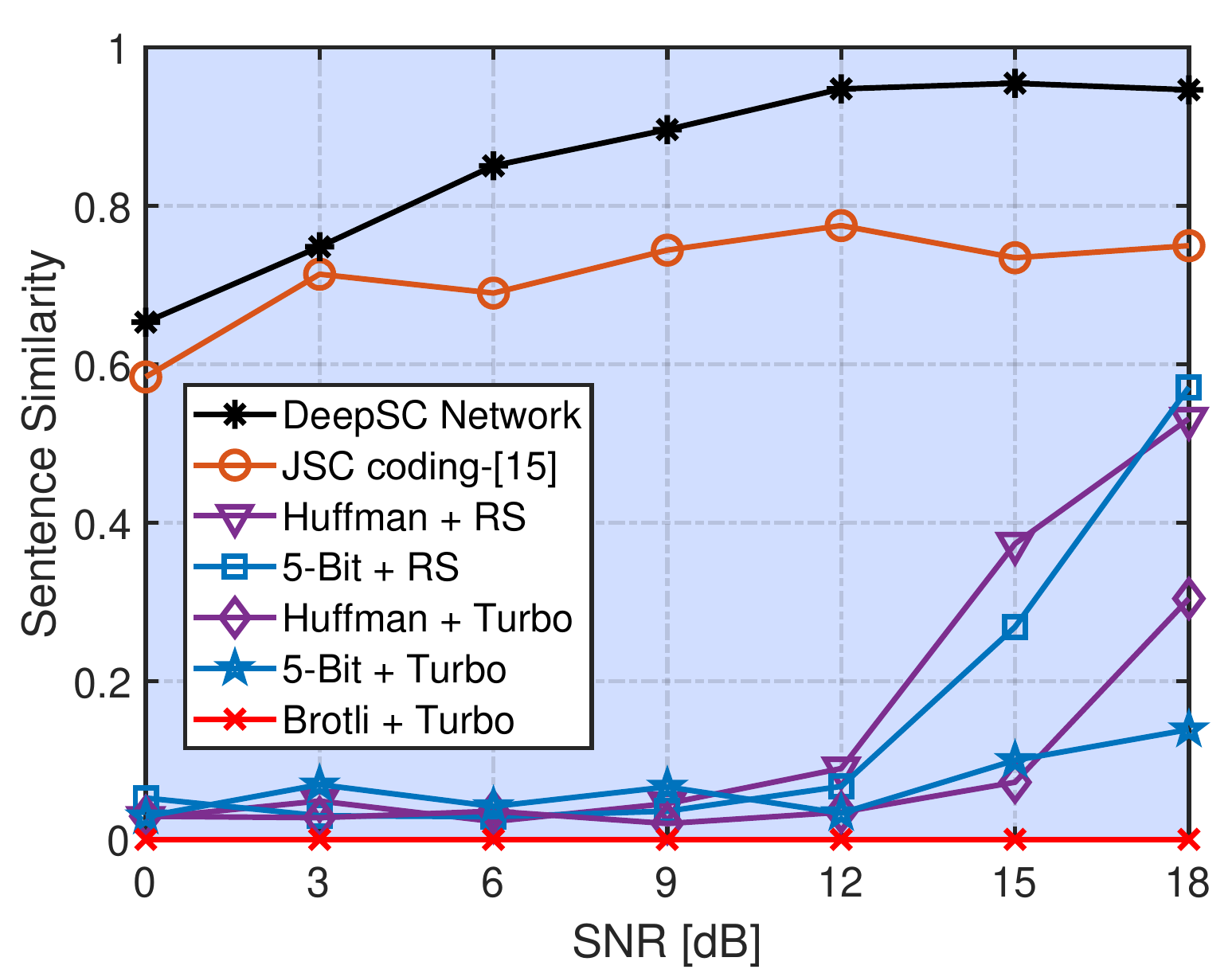}
			}
		\end{center}
		\vspace*{-3mm}
		\captionsetup{font={footnotesize}, name={Fig.},labelsep=period}
		\caption{(a)~The transformer-based semantic communication architecture proposed in \cite{SC_DL_1}; and (b)~Sentence similarity of various schemes vs. SNR for the same total number of transmitted symbols over the Rayleigh fading channel quoted from \cite{SC_DL_1}.}
		\label{SC} 
		\vspace*{-4mm}
	\end{figure*}
	
	\subsection{Semantic Communication}\label{S3.4} 
	Our communication networks have been traditionally conceived and designed as bit pipes; that is, the goal has been to deliver as many bits as possible with the highest reliability. Current communication networks do not take into account the meaning or the purpose of the delivered bits, whose interpretation and processing have been left to higher layers. To meet the requirements of 6G wireless networks, however, it is important to propose more efficient information acquisition and delivery methods. The recently growing trend of semantic communication aims at accurately recovering the statistical structure of the underlying source signals and designing the communication system in an end-to-end fashion, similarly to joint source and channel (JSC) coding by taking the source semantics into account \cite{SC-CNN,SC_DL_1,SC-LSTM}. 
	Fig.~\ref{SC}\,(a) shows the general framework of a semantic communication model, where the transmitter includes a semantic encoder and a semantic-aware JSC encoder, and the receiver includes a semantic-aware JSC decoder and a semantic decoder. In general, the transmitter can perform semantic encoding on the source according to the knowledge library for obtaining highly compressed abstract semantics, followed by JSC encoder and subsequent baseband signal processing. The receiver follows the reverse steps of the transmitter, where a JSC decoder is followed by a semantic decoder based on some knowledge library. Alternatively, the semantic and JSC encoder/decoder operations can be considered into single module as in \cite{SC-CNN}.
	
	Semantic communication is particularly effective for complex information sources, such as text, speech, image, or video, where the reconstruction quality depends on the source semantics, and is often difficult to measure through traditional measures of bit error rate or mean square error.
	In \cite{SC-LSTM}, the authors proposed a LSTM-based model to extract the semantic information of sentences through JSC coding for text transmission. However, due to the lack of a separate semantic coding module, JSC coding can only implicitly utilize the semantic information, which has difficulty to represent specific semantics. Instead, the transformer can extract correlations between different words to form highly abstract semantics. Inspired by this benefit, a DL-enabled semantic communication (DeepSC) scheme was proposed in \cite{SC_DL_1}, where a separate semantic coding network is utilized to better extract accurate semantic information.
	As shown in Fig.~\ref{SC}\,(a), a transformer encoder is utilized as the semantic encoder and a MLP is used as the JSC encoder. The Rayleigh fading channel is interpreted as an untrainable layer in the model. Correspondingly, the receiver consists of a MLP-based JSC decoder followed by a transformer decoder for text reconstruction. The whole network is trained in an end-to-end fashion to simultaneously minimize the sentence similarity and maximize mutual information. Fig.~\ref{SC}\,(b) compares the performance of DeepSC \cite{SC_DL_1} in transmitting text over a Rayleigh fading channel with the following benchmarks: Huffman code followed by Reed-Solomon (RS) coding and 64-quadrature amplitude modulation (QAM), fixed-length code (5-bit) followed by RS coding and 64-QAM, Huffman code followed by Turbo coding and 64-QAM, 5-bit code followed by Turbo coding and 128-QAM, Brotli code followed by Turbo coding and 8-QAM, and the JSC coding approach of \cite{SC-LSTM}. The simulation results demonstrate that thanks to the powerful transformer architecture, the sentence similarity performance of DeepSC \cite{SC_DL_1} far outperforms the alternatives based on separate compression followed by channel coding, as well as the JSC coding approach \cite{SC-LSTM}. 
	Hence, we foresee that semantic-aided communication is an important challenge, where the transformer architecture is likely to have an impact on future communication systems by more effectively learning and adapting to the statistics of complex signals, such as text, image, or video. The systematic design of transformers may also allow designing a common architecture for the communication of multiple modalities.
	
	While we have considered a simple single-input single-output channel in the example in Fig.~\ref{SC}, 6G communication networks will need to combine semantic communication with massive MIMO and other core communication tools and techniques, as illustrated in Fig.~\ref{MIMO_framework}\,(b). This will require jointly optimizing these modules in an end-to-end fashion. One of the challenges facing semantic communications is to achieve the potential gains from the joint processing of source and channel coding with other components while retaining the low-complexity and modular network architecture.
	
	
	\section{Challenges and Open Issues}\label{III}
	
	We hope that the above examples have convinced the readers of the significant potential of the transformer architecture for future 6G intelligent network design. In addition to these examples, we expect that the transformers will find applications in waveform design, channel modeling and generation, signal detection, as well as more advanced sensing techniques exploiting other complex information sources such as LiDAR or cameras. We would like to highlight that the transformer architecture was invented only in 2017. Although it has received significant attention in the last years thanks to its superior performance, the research on transformer-based communication system design is still in its infancy, and many key issues are still open. In this section, we discuss several potential directions for future study.

	\textbf{Network Efficiency and Generalization}: Network efficiency and generalization problems have been widely discussed in the context of DL-assisted communication systems. They are particularly severe in transformer architectures and may limit their further adoption. An important limitation of the transformer architecture is the high computation and memory complexity, mainly caused by the self-attention module. Recently, various model variants have been proposed to improve computational and memory efficiency, such as sparse attention, linearized attention, low-rank self-attention\cite{Transformer-Survey}. Moreover, since the transformer makes few assumptions on the structural bias of the input data, the network cannot perform real-time parameter retraining to overcome poor generalization. Transfer learning and introducing structural biases or regularization can be considered to address this issue. However, existing works mainly focus on CV and NLP applications. Hence, to successfully apply the transformer architecture to 6G networks, an important research challenge is to optimize their computational efficiency and generalization capabilities by developing effective and efficient transformer architectures targeting wireless applications.
	
		
	\textbf{Efficient Information Injection}: In NLP, the text is divided into words, and a word embedding is used to feed each word to the transformer network. Similarly, in CV, each image is divided into patches, and the sequences of linear embedding of these patches are fed as input to a transformer. Similar techniques can be used for the semantic communication of text and image sources; however, for the physical layer design, the input is mainly based on CSI, which commonly has four dimensions: time-space-frequency-user. In this article, the inputs of channel estimation and CSI feedback take self-attention on the frequency domain of CSI, while in the hybrid beamforming, the frequency and user domains are used jointly. Therefore, how to efficiently feed the underlying input, which can include the source signal, CSI tensor, location, traffic and environment information, input the transformer architecture is one of the topics to be investigated for wireless applications.
	
	\textbf{Combination with Model-Driven DL}: Model-driven DL methods introduce learnable parameters while retaining the model assumptions and the often-used iterative optimization of the model parameters, such as LAMP \cite{LAMP}. This usually results in faster convergence and requires a smaller dataset. However, the performance can deteriorate severely when the underlying model is inaccurate, e.g., in the low SNR regime, or with non-Gaussian noise. In order to further improve the robustness of model-driven DL algorithms, there has been some work on employing deep neural networks, such as CNNs, to replace the original MMSE denoiser to refine the output. Compared with other DL architectures, the transformer has a distinguished feature extraction capability from data. Therefore, integrating the transformer architecture into the model-driven framework is a promising approach to further mitigate the performance degradation caused by model inaccuracies.
	
	\textbf{Parallel Communication Sequential Tasks}: Transformer architecture is significantly superior to conventional RNN models in sequential tasks, thanks to its ability to use self-attention for capturing various long-term temporal correlations in parallel and to learn a better representation for predicting the next state. Thus, transformers are also expected to be successful in communication tasks involving temporal sequences, such as channel prediction and beam tracking. 
	
	\section{Conclusions}\label{S5}
	
	In this article, we have presented the transformer architecture and provided examples to highlight its potential benefits in addressing various challenges for 6G intelligent networks. We have considered the applications of transformers from massive MIMO processing to semantic communication, and provided concrete examples to show their competitive performance compared to the other classical as well as recently proposed DL-based models, hence demonstrating their significant potential for designing AI-native future communication systems. Potential research directions have also been identified to encourage efforts by the research community to further develop a transformer-based 6G intelligent network paradigm.
	
	\section{Acknowledgment}\label{S6}
	
	The work of Z. Gao was supported in part by the Natural Science Foundation of China (NSFC) under Grant 62071044 and Grant 62088101, in part by the Shandong Province Natural Science Foundation under Grant ZR2022YQ62. The work of H. V. Poor was supported in part by the U.S. National Science Foundation under Grant CNS-2128448, and in part by a grant from the C3.ai Digital Transformation Institute.

\section*{Biographies}
\textsc{Yang Wang} is currently pursuing the M.S. degree with the School of Information and Electronics, Beijing Institute of Technology, China. His research interests include UAV communications, deep learning for wireless communications, and RIS.

\vspace{8 pt}

\textsc{Zhen Gao} is an Associate Professor with the Beijing Institute of Technology. His research interests are in wireless communications, with a focus on multi-carrier modulations, multiple antenna systems, and sparse signal processing. He is the corresponding author of this article.

\vspace{8 pt}

\textsc{Dezhi Zheng} is a Professor with the Beijing Institute of Technology. His research interests include sensor technology, signal detection, and processing technology.

\vspace{8 pt}

\textsc{Sheng Chen} (Fellow, IEEE) is a Professor with the School of Electronics and Computer Science, the University of Southampton, U.K. Professor Chen's research interests include neural network and machine learning, adaptive signal processing, and wireless communications. Dr. Chen is a Fellow of the United Kingdom Royal Academy of Engineering, a Fellow of Asia-Pacific Artificial Intelligence Association and a Fellow of IET.

\vspace{8 pt}

\textsc{Deniz Gündüz} (Fellow, IEEE) is a Professor of Information Processing at Imperial College London, and serves as the deputy head of the Intelligent Systems and Networks Group. His research interests lie in the areas of communications, information theory, machine learning and privacy. Dr. Gündüz is a Fellow of the IEEE, and a Distinguished Lecturer for the IEEE Information Theory Society (2020-22).

\vspace{8 pt}

\textsc{H. Vincent Poor} (Life Fellow, IEEE) is the Michael Henry Strater University Professor at Princeton University. His research interests are in the areas of information theory, machine learning and network science, and their applications in wireless networks, energy systems and related fields. Dr. Poor is a member of the National Academy of Engineering and the National Academy of Sciences and is a foreign member of the Chinese Academy of Sciences, the Royal Society, and other national and international academies. He received the IEEE Alexander Graham Bell Medal in 2017.

\end{document}